\documentclass[reprint,prc,
 amsmath,amssymb,
 aps,
 lengthcheck,
floatfix,
]{revtex4}
\usepackage{balance}
\usepackage{graphicx}
\usepackage{dcolumn}
\usepackage{bm}
\usepackage{hyperref}
\usepackage[utf8]{inputenc}
\usepackage{amsmath}
\usepackage{multirow}
\usepackage{soul}
\usepackage{float}
\usepackage{graphicx}
\usepackage{times}
\usepackage[normalem]{ulem}
\usepackage{color}
\usepackage{cellspace}
\usepackage{lipsum}

\newcommand{\be}{\begin{equation}}
\newcommand{\ee}{\end{equation}}
\newcommand{\bea}{\begin{eqnarray}}
\newcommand{\eea}{\end{eqnarray}}

\newcommand {\nonu}{\nonumber}

\newcommand{\comment}[1]{}
\renewcommand\sout{\bgroup \color{red} \ULdepth=-.5ex \ULset}
\def\simge{\mathrel{\rlap{\raise 0.511ex
     \hbox{$>$}}{\lower 0.511ex \hbox{$\sim$}}}}
\def\simle{\mathrel{\rlap{\raise 0.511ex
      \hbox{$<$}}{\lower 0.511ex \hbox{$\sim$}}}}

\begin{document}


\title{Implications of comprehensive nuclear and astrophysics data\\ on the equations of state of neutron star matter}

\author{Sk Md Adil Imam$^{1,2}$}
\email{skmdadilimam@gmail.com}
\author{Tuhin \surname{Malik}$^3$}
\email{tm@uc.pt}
\author{Constança \surname{Providência}$^3$}
\email{cp@uc.pt}
\author{B. K. \surname{Agrawal}$^{1,2,3}$}
\email{bijay.agrawal@saha.ac.in}

\affiliation{$^1$Saha Institute of Nuclear Physics, 1/AF 
Bidhannagar, Kolkata 700064, India.}  
\affiliation{$^2$Homi Bhabha National Institute, Anushakti Nagar, Mumbai 400094, India.}  

\affiliation{$^3$CFisUC, Department of Physics, University of Coimbra,
3004-516 Coimbra, Portugal}

\date{\today}
\begin{abstract} 
The equations of state (EoSs) governing neutron star (NS) matter obtained for both non-relativistic and relativistic mean-field models are systematically confronted with a diverse set of terrestrial data and  astrophysical observations within the Bayesian framework. The terrestrial data, spans  from bulk properties of finite nuclei to the heavy-ion collisions, constrain the symmetric nuclear matter EoS and the symmetry energy up to twice the saturation density ($\rho_0$= 0.16 fm$^{-3}$). The astrophysical observations encompass the NS radius, the tidal deformability, and the lower bound on maximum mass. Three distinct posterior distributions of EoSs are generated by gradually updating the priors with different constraints: (i) only the maximum NS mass, (ii) incorporating additional terrestrial data, (iii) combining both the terrestrial data and astrophysical observations. These EoS distributions are then compared using the Kullback-Liebler divergence which highlights the significant constraints imposed on the EoSs by the currently available lower bound of NS maximum mass and terrestrial data. The remaining astrophysical observations marginally refine the EoS within the density range $\sim$ 2-3$\rho_0$. It is observed that the relativistic mean field model yields stiffer EoS around the saturation density, but predict smaller values of the speed of sound and proton fraction in the interior of massive stars.
\end{abstract}
\maketitle

\section{Introduction}
Determining the unified equation of state (EoS) accurately requires inputs from both nuclear physics experiments and astrophysical observations. The experimental data on heavy-ion collisions (HICs) and bulk properties of finite nuclei  confine the behavior of the EoS at low densities ($\rho \leq 2\rho_0$). These data specifically constrain the EoS for symmetric nuclear matter (SNM) and the density-dependent symmetry energy, crucial elements for determining the EoS of neutron star matter \cite{Lattimer:2000nx,Lattimer_rev,Fattoyev:2010mx,Tsang:2012se,Fortin16,Margueron:2017lup,Fattoyev2018a,Greif_2020,Reed:2023cap}.
The astrophysical observations such as neutron star radius, tidal deformability, and maximum mass further constrain the EoS at high densities~\cite{Raaijmakers:2021uju,Annala2022,Ghosh_mpc,Huang:2023grj,Traversi:2020aaa,Most:2018hfd}.
The radius and tidal deformability parameter of neutron stars  are sensitive to the EoS at supra-saturation densities. They have been deduced from the gravitational wave events GW170817 and GW190425 observed using the Advanced-LIGO~\cite{LIGOScientific:2014pky} and Advanced-Virgo detectors~\cite{VIRGO:2014yos}. In particular, the gravitational waves associated to the event GW170817 stemmed from a binary neutron star (BNS) merger with component masses ranging from 1.17 to 1.6 M$_\odot$~\cite{Abbott18a,Abbott2019} has sparked numerous theoretical investigations into the neutron star properties \cite{GW170817, Abbot2018, De18, Fattoyev2018a, Landry2019, Piekarewicz2019, Abbott2020, Dietrich:2020lps}. 
Future observations of coalescing BNS events by detectors like LIGO-Virgo-KAGRA, Einstein Telescope~\cite{Punturo:2010zz}, and Cosmic Explorer~\cite{Reitze:2019iox} are likely to occur more frequently, enabling a more precise determination of the EoS \cite{PhysRevD.108.122006,Huxford:2023qne,Rose:2023uui,walker2024precision,Pradhan_2023,pradhan2023cost}. 
The Neutron star Interior Composition Explorer (NICER) offers complementary information on the  NS properties ~\cite{Watts2016, Psaltis2014}. 

The nuclear matter parameters are the key quantities that determine the EoS. The combined constraints from the bulk properties of finite nuclei and NS observable has been employed to constrain the nuclear matter parameters which are the expansion coefficients in the Taylor model for the EoS \cite{Zhang_2018}. The finite nuclei constraints are encoded through a few low order nuclear matter parameters treating them independent of each other. The NS observable considered are  the radius, 10.62 km $<R_{1.4}<$ 12.83 km \cite{Lattimer:2014sga}, the dimensionless tidal deformability  $\Lambda_{1.4}\leq$800   \cite{GW170817,LIGOScientific:2017ync} and lower limit of the maximum mass imposed by the NS PSR J0348
- 0432 with a mass 2.01 $\pm$ 0.04 M$_\odot$ \cite{Antoniadis:2013pzd,lattimer2014constraints}. It is demonstrated that these constraints limit the multidimensional space of few higher order nuclear matter parameters. It may be emphasized that the astrophysical inputs employed in Ref.\cite{Zhang_2018} have undergone significant revision. Two different groups of NICER have reported neutron star's mass and radius simultaneously for PSR J0030+0451 with radius R $=13.02^{+1.24}_{-1.06}$km for mass M $=1.44^{+0.15}_{-0.14}$ M$_\odot$~\cite{miller2019} and R $=12.71^{+1.14}_{-1.19}$ km for M $=1.34^{+0.15}_{-0.16}$ M$_\odot$~\cite{riley2019}. Recently, another (heavier) pulsar PSR J0740+6620, R $=13.7^{+2.6}_{-1.5}$ km with M $=2.08 \pm 0.07$ M$_\odot$~\cite{miller2021} and R $=12.39^{+1.30}_{-0.98}$ km with M $=2.072^{+0.067}_{-0.066}$ M$_\odot$ \cite{riley2021} have been reported. The improvement in the precision of astrophysical observation and the terrestrial data has triggered the search for the universal EoS . 

A Bayesian inference technique has been employed \cite{Huth:2021bsp} to combine data from astrophysical multi-messenger observations of neutron stars ~\cite{GW170817,Abbott:2020uma,Monitor:2017mdv,Coughlin:2017ydf,miller2019,riley2019,miller2021,riley2021,Dietrich:2020lps}, high-energy heavy-ion collisions involving gold nuclei ~\cite{LeFevre:2015paj,Russotto16} and microscopic nuclear theory calculations ~\cite{Hebeler:2013nza,Tews:2012fj,Lynn:2016, Drischler:2017wtt, Dris20GPB, Huth_2021} to constrain the EoS of dense matter. The microscopic input was extended up to 1.5 times the nuclear saturation density ($\rho_0$) for the $\beta$-EoS derived from chiral effective field theory ($\chi$EFT). The nuclear experimental data includes the symmetry energy and symmetric nuclear matter pressure extracted from the FOPI ~\cite{LeFevre:2015paj} and the ASY-EoS ~\cite{Russotto16} experiments, respectively. The astrophysical data included mass measurements of massive neutron stars PSR J0348+0432 \cite{Antoniadis:2013pzd} and PSR J1614-2230 \cite{Arzoumanian:2017puf}. Constraints on the maximum mass of neutron stars were also derived from the binary neutron-star collision GW170817~\cite{Margalit:2017dij, Rezzolla:2017aly}. The NS radii of PSR J0030+0451 and PSR J0740+6620 utilizing NICER and XMM-Newton data~\cite{miller2019, miller2021, riley2021} was also included. A Bayesian inference technique was employed to analyze the GW information from the GW170817~\cite{GW170817} and GW190425~\cite{Abbott:2020uma} events, aligning observed GW data with theoretical models reliant on neutron-star properties. The analysis relied on a GW model~\cite{Dietrich:2019kaq}, an enhanced version of the primary waveform model utilized by the LIGO/Virgo Collaboration in the exploration of GW170817~\cite{Abbott18a} and GW190425~\cite{Abbott:2020uma}. Information pertaining to the kilonova AT2017gfo~\cite{Monitor:2017mdv}, linked with the GW signal, was also incorporated. Electromagnetic observations underwent comprehensive radiative transfer simulations~\cite{Bulla:2019muo} to extract insights from observed light curves and spectra~\cite{Coughlin:2017ydf}. However, the EoS employed in this analysis extended the $\chi$EFT calculation to $\beta$-equilibrium with a crust \cite{Tews:2016ofv}, while beyond this density range, a collection of six randomly distributed points within the speed of sound (c$_s$) plane at baryon densities spanning 1.5 to 12$\rho_0$ were sampled, enforcing 0 $\leq$c$_s \leq$ 1 at each point.

Of late \cite{Tsang2024}, the EoS has been constrained using a diverse set of data from nuclear experiments and astrophysical observations. The nuclear inputs consist of experimental data on bulk properties of finite nuclei and HIC and astrophysical observations on NS properties. The properties of finite nuclei and HIC are encoded through the symmetry energy, pressure for symmetry energy and SNM spanning the density up to 2$\rho_0$. The experimental data on nuclear masses, iso-vector giant dipole resonance, neutron skin thickness and isobaric analog states constrain the symmetry energy and the symmetry energy pressure at sub-saturation densities \cite{brown2013constraints,decoding2022,kortelainen2012nuclear,danielewicz2017symmetry,zhang2015electric,adhikari2021accurate,reed2021implications}. The HIC data from isospin diffusion, neutron-proton ratios, charged pion spectra are incorporated to constrain the behavior of SNM and symmetry energy from sub-saturation to supra-saturation densities \cite{tsang2009constraints,decoding2022,morfouace2019constraining,estee2021probing,cozma2018feasibility,RUSSOTTO2011471,Russotto16,Danielewicz:2002pu,LeFevre:2015paj}. The recent astrophysical observations for the radius of a NS with mass $\sim$ 1.4 and 2.1 M$_\odot$  \cite{riley2019,miller2019,Antoniadis:2013pzd} and the dimensionless tidal deformability for 1.4 M$_\odot$ NS\cite{Abbot2018} have also been considered. These data are used in a Bayesian framework to constrain the EoS obtained using Taylor expansion. 

We use a Bayesian framework for the systematic analysis of the implications of the diverse set of data \cite{Tsang2024} in constraining the EoS of NS matter obtained within a Skyrme based non-relativistic mean field and non-linear variant of relativistic mean field (RMF) models. Our analysis involves three distinct sets of calculations : \\
(i) Priors updated with the NS maximum mass.\\
(ii) Further inclusion of 12 experimental data.\\
(iii) Inclusion of the experimental data as well as astrophysical observations corresponding to the NS properties.

The constraints from astrophysical observations are incorporated directly by employing the mass-radius posterior distribution for PSR J0030+0451\cite{riley2019,miller2019} and PSR J0740+6620\cite{miller2021, riley2021} and posterior distribution for dimensionless tidal deformability for binary neutron star components from the GW170817 event \cite{Abbott18a}.



This paper is organised as follows. In Sec.\ref{meth} a brief description of mean field models and Bayesian framework are presented. Sec.\ref{data} summarizes the experimental data and astrophysical observations considered. The results are discussed in Sec.\ref{res}. Finally, the conclusions are drawn in Sec.\ref{conc}.

\section{Methodology}\label{meth}
We use non-relativistic and the relativistic mean field models for the EoSs. The non-relativistic mean field models are based on Skyrme type effective interaction. The relativistic mean field models are derived from the effective Lagrangian which describes the interactions of nucleons through scalar-isoscalar meson $\sigma$, vector-isoscalar meson $\omega$ and vector-isovector meson $\varrho$  which also includes non-linear self interaction terms for $\sigma$ mesons and mixed interaction term involving $\omega$ and $\rho$ mesons. Once we have these EoSs we can calculate the energy per nucleon of the SNM, e$_{\rm SNM}(\rho)$ and symmetry energy, e$_{\rm sym}(\rho)$ and use them to obtain the pressure of the symmetric nuclear matter (P$_{\rm SNM}$) and the symmetry energy pressure (P$_{\rm sym}$). Empirical values of these quantities are deduced from experimental data on finite nuclei and HIC which are employed to constrain the required EoS.

\subsection{SKYRME EoS}

The Hamiltonian density for the Skyrme type effective interaction contributing to uniform nuclear matter can be written as
\begin{equation}
	\label{Hden}
	{\cal H} = {\cal K} + {\cal H}_0  +{\cal H}_3+{\cal H}_{\rm eff}
\end{equation}
where, ${\cal K} = \frac{\hbar^2}{2m}\tau$ is the kinetic energy term,
${\cal H}_0$ is the zero-range  term, ${\cal H}_3$ the density dependent
term, ${\cal H}_{\rm eff}$ an effective-mass term. For the Skyrme interaction  we have,

\begin{equation}
	{\cal H}_0 = \frac{1}{4}t_0\left [(2 + x_0)\rho^2 - (2x_0 + 1)(\rho_p^2 + \rho_n^2)\right ],
\end{equation}
\begin{equation}
	{\cal H}_3 = \frac{1}{24}t_3\rho^\sigma\left [(2 + x_3)\rho^2 - (2x_3 + 1)(\rho_p^2 + \rho_n^2)\right ],
\end{equation}
\bea
\nonumber
{\cal H}_{\rm eff} &=& \frac{1}{8} \left [t_1 (2 + x_1) + t_2 (2 + x_2)\right ]
\tau\rho \\
&+&\frac{1}{8}\left[ t_2(2x_2+1) - t_1(2x_1+1)\right] (\tau_p\rho_p + \tau_n\rho_n), 
\eea

Here, $ \rho = \rho_p + \rho_n,$ and  $\tau = \tau_p + \tau_n,$ {\rm }
are the particle number density and kinetic energy density with $p$ and $n$ denoting the
protons and neutrons, respectively.
\bea
\tau_q &=& \frac{\gamma}{(2\pi)^3}\int_{0}^{k_{f_q}}k^2d^3k\nonu \\
&=&\frac{3}{5}(3\pi^2)^{2/3}{\rho_q}^{5/3}
\eea
In the above q denotes p, n and $\gamma$=2 for spin degeneracy. Energy per nucleon for symmetric nuclear matter (i.e. $\rho_p = \rho_n = \frac{\rho}{2}$) is given as,
\begin{align}
	e_{\rm SNM}(\rho) &=& a_k\rho^{\frac{2}{3}} + \frac{3t_0}{8}\rho + \frac{3}{10}\theta_{\rm SNM}(\frac{3\pi^2}{2})^\frac{2}{3}\rho^\frac{5}{3} + \frac{t_3}{16}\rho^{\sigma+1}\label{eq:SNM}
\end{align}
Where a$_k$ = $\frac{3\hbar^2}{10m}(\frac{3\pi^2}{2})^\frac{2}{3}$ and $\theta_{\rm SNM} = \frac{1}{8}(3t_1 + (5+4x_2)t_2)$.
We can express the Skyrme parameters in the Eq.(\ref{eq:SNM}) in terms of the nuclear matter parameters such as  binding energy $e_0$, incompressibility K$_0$  and effective mass m$^*$ evaluated at the saturation density $\rho_0$, as \cite{Chabanat97,Margueron02,Gomez92,Agrawal:2005ix}
\begin{eqnarray}
	\label{t0}
	t_0 &=& (\frac{8}{\rho_0})*(\frac{f_1}{f_2} + f_3) \\
	\sigma & = & \frac{f_5}{f_6} \\
	t_3 &=& (\frac{16}{\rho_0^{\sigma+1}})(\frac{f_4}{f_2})\\ 
	\label{theta_snm}
	\theta_{SNM} &=&\frac{\hbar^2}{m\rho_0}(\frac{m}{m^*}-1)\\ \nonumber
\end{eqnarray}
Where,
\begin{eqnarray}
	\nonumber
	f_1 & = & (e_0 + (\frac{2m}{m^*}-3)(\frac{\hbar^2}{10m})k_f^2)\\ \nonumber
	& &(\frac{K_0}{27} -(1-\frac{6m^*}{5m})(\frac{\hbar^2}{9m^*})k_f^2)\\
	\nonumber
	f_2 & = & e_0+\frac{K_0}{9}-(\frac{4m}{3m^*}-1)(\frac{\hbar^2}{10m})k_f^2 \\
	\nonumber
	f_3 & = & (1-\frac{5m}{3m^*})(\frac{\hbar^2}{10m})k_f^2\\
	\nonumber
	f_4 &=& (e_0 + (\frac{2m}{m^*}-3)(\frac{\hbar^2}{10m})k_f^2)^2 \\
	\nonumber
	f_5 & = & -e_0 - \frac{K_0}{9} + (\frac{4m}{3m^*}-1)(\frac{\hbar^2}{10m})k_f^2 \\
	\nonumber
	f_6 & = & e_0 + (\frac{2m}{m^*}-3)(\frac{\hbar^2}{10m})k_f^2 \\ 
	\nonumber
\end{eqnarray}

The symmetry energy can be expressed as,
\bea
e_{sym}(\rho) &=& a_{k1}\rho^{\frac{2}{3}} + a_0\rho + a_{sym}\rho^{\frac{5}{3}} +a_3\rho^{\sigma+1}\label{eq:sym}
\eea
Here a$_{k1}=\frac{5}{9}a_k$. The values of a$_0$, a$_{sym}$ and a$_3$ can be computed using the nuclear matter parameters J$_0$, L$_0$ and K$_{sym0}$ as
\bea
\nonu
a_3 &=&\frac{K_{\rm sym0} -5(L_0-3J_0)-3a_{k1}\rho_0^{\frac{2}{3}}}{3\sigma(3\sigma-2)\rho_0^{(\sigma+1)}}\\ \nonu 
a_{sym} &=&\frac{1}{2}((L_0-3J_0)-3\sigma a_3\rho_0^{\sigma+1} + a_{k1}\rho_0^\frac{2}{3})\rho_0^{-\frac{5}{3}}\\ \nonu
a_{0} &=& \frac{J_0}{\rho_0} -a_{k1}\rho_0^{-\frac{1}{3}} - a_{sym}\rho_0^{\frac{2}{3}} -a_3\rho_0^{\sigma}
\eea
Once a$_0$, a$_{sym}$ and a$_3$ are known the Skyrme parameters x$_0$, $\theta_{sym}$ and x$_3$ can be computed as
\bea
\label{x0}
x_0 &=&-\frac{1}{2}(\frac{8a_0}{t_0} + 1) \\ 
\theta_{sym} &=& -24a_{sym}(\frac{3\pi^2}{2})^{-\frac{2}{3}}\\
\label{x3}
x_3 &=&-\frac{1}{2}(\frac{48a_3}{t_3} + 1)
\eea\label{eq: sym_param}
With $\theta_{\rm sym} = 3t_1x_1 - t_2(5 + 4x_2)$. For a given set of NMPs the Skyrme parameters can be computed using Eqs.(\ref{t0}-\ref{theta_snm}) and Eqs.(\ref{x0}-\ref{x3}).
Once these Skyrme parameters are known the energy density for a given $\rho$ and asymmetry, $\delta =\frac{\rho_n-\rho_p}{\rho}$ can be computed using e$_{\rm SNM}(\rho)$ and e$_{\rm sym}(\rho)$.
\vspace{1cm}
\begin{table*}
\caption{\label{tab1}The empirical values  of symmetry energy (e$_{\rm sym}$), symmetry energy pressure (P$_{\rm sym}$) and symmetric nuclear matter pressure (P$_{\rm SNM}$) from experimental data on the bulk properties of finite nuclei and HIC. The astrophysical observational constraints of radii and tidal deformability of neutron star. See Ref\cite{Tsang2024} for details.}

\begin{tabular*}{\textwidth}{@{\extracolsep{\fill}}lcccc}
\hline
Symmetric matter  &&&& \\
Constraints & $n$ (fm$^{-3}$) & P$_\text{SNM}$ (MeV/fm$^3$)&  & Ref. \\
\hline
HIC(DLL)  & 0.32             & $10.1\pm3.0$                 &                      &\cite{Danielewicz:2002pu} \\
HIC(FOPI)  & 0.32                & $10.3\pm2.8$                 &                      &\cite{LeFevre:2015paj} \\ \\
\hline
\hline
Asymmetric matter  &&&& \\
Constraints & $n$ (fm$^{-3}$) & S($n$) (MeV)      & P$_\text{\rm sym}$ (MeV/fm$^3$) & Ref. \\ 
\hline
Nuclear structure &&&&\\
$\alpha_D$    & 0.05           & $15.9\pm1.0$ &                                        &~\cite{zhang2015electric}        \\
PREX-II       & 0.11             &              & $2.38\pm0.75$           &~\cite{adhikari2021accurate,reed2021implications,decoding2022}\\
\\
Nuclear masses &&&&\\ 
Mass(Skyrme)  & 0.101          & $24.7\pm0.8$ &                                          &~\cite{brown2013constraints,decoding2022}
\\
Mass(DFT) &  0.115             & $25.4\pm1.1$ &                                             &~\cite{kortelainen2012nuclear,decoding2022}  \\
IAS           & 0.106           & $25.5\pm1.1$ &                                             &~\cite{danielewicz2017symmetry,decoding2022}     \\
\\
Heavy-ion collisions \\ 
HIC(Isodiff)  & 0.035       & $10.3\pm1.0$ &                                             &~\cite{tsang2009constraints,decoding2022}    \\
HIC(n/p ratio) & 0.069\          & $16.8\pm1.2$ &                                         &~\cite{morfouace2019constraining,decoding2022}\\
HIC($\pi$)    & 0.232           & $52\pm13$    & $10.9\pm8.7$                         &~\cite{estee2021probing,decoding2022}    \\

HIC(n/p flow) & 0.240           &              & $12.1\pm8.4$          &~\cite{cozma2018feasibility,RUSSOTTO2011471,Russotto16,decoding2022} \\
\\
\hline
\hline
\end{tabular*}
\begin{tabular*}{\textwidth}{@{\extracolsep{\fill}}lcccc}
Astronomical    &&&& \\
Constraints & $M_\odot$ & R (km) & $\Lambda_{1.36}$ & Ref.\\
\hline 
LIGO~\footnote{LVK collaboration,~\href{https://dcc.ligo.org/LIGO-P1800115/public}{https://dcc.ligo.org/LIGO-P1800115/public}} & 1.36 & & $300_{-230}^{+420}$ &~\cite{Abbott18a} \\ \vspace{0.2mm}
*Riley ~PSR J0030+0451~\footnote{~\href{https://zenodo.org/records/8239000}{https://zenodo.org/records/8239000}}& 1.34 & $12.71^{+1.14}_{-1.19}$ & &~\cite{riley2019} \\\vspace{0.2mm}

*Miller PSR J0030+0451~\footnote{~\href{https://zenodo.org/record/3473466\#.XrOt1nWlxBc}{https://zenodo.org/record/3473466\#.XrOt1nWlxBc}}& 1.44 & $13.02^{+1.24}_{-1.06}$ & &~\cite{miller2019} \\\vspace{0.2mm}
*Riley ~PSR J0740+6620~\footnote{~\href{https://zenodo.org/records/4697625}{https://zenodo.org/records/4697625}} & 2.07 & 12.39$^{+1.30}_{-0.98}$ & &~\cite{riley2021}\\\vspace{0.2mm}
*Miller PSR J0740+6620~\footnote{~\href{https://zenodo.org/records/4670689}{https://zenodo.org/records/4670689}} & 2.08 & $13.7~^{+2.6}_{-1.5}$ & &~\cite{miller2021}
 \\
\hline
\end{tabular*}
\end{table*}
\subsection{RMF EoS} \label{RMF}
We also create a collection of EoSs within a model of nuclear matter that is based on a relativistic field theory approach. The nuclear interaction between nucleons is represented by the exchange of the scalar-isoscalar meson $\sigma$, the vector-isoscalar meson $\omega$, and the vector-isovector meson $\varrho$. 
 
The Lagrangian density is given by \citep{Fattoyev:2010mx,Dutra:2014qga,Malik:2023mnx}
        \begin{equation}
          \mathcal{L}=   \mathcal{L}_N+ \mathcal{L}_M + \mathcal{L}_{NL} +\mathcal{L}_{leptons}
        \end{equation} 
where $$\mathcal{L}_{N} = \bar{\Psi}\Big[\gamma^{\mu}\left(i \partial_{\mu}-g_{\omega} \omega_\mu - \frac{1}{2}g_{\varrho} {\boldsymbol{t}} \cdot \boldsymbol{\varrho}_{\mu}\right) - \left(m_N - g_{\sigma} \sigma\right)\Big] \Psi$$ denotes the Dirac equation for the nucleon doublet (neutron and proton) with bare mass $m_N$, $\Psi$ is a Dirac spinor, $\gamma^\mu $ are the Dirac matrices, and $\boldsymbol{t}$ is the isospin operator. $\mathcal{L}_{M}$ is the Lagrangian density for  the mesons, given by
\begin{eqnarray}
\mathcal{L}_{M}  &=& \frac{1}{2}\left[\partial_{\mu} \sigma \partial^{\mu} \sigma-m_{\sigma}^{2} \sigma^{2} \right] - \frac{1}{4} F_{\mu \nu}^{(\omega)} F^{(\omega) \mu \nu} + \frac{1}{2}m_{\omega}^{2} \omega_{\mu} \omega^{\mu}   \nonumber \\
  &-& \frac{1}{4} \boldsymbol{F}_{\mu \nu}^{(\varrho)} \cdot \boldsymbol{F}^{(\varrho) \mu \nu} + \frac{1}{2} m_{\varrho}^{2} \boldsymbol{\varrho}_{\mu} \cdot \boldsymbol{\varrho}^{\mu} \nonumber
\end{eqnarray}
where $F^{(\omega, \varrho)\mu \nu} = \partial^ \mu A^{(\omega, \varrho)\nu} -\partial^ \nu A^{(\omega, \varrho) \mu}$ are the vector meson  tensors, and
\begin{eqnarray}
\mathcal{L}_{NL}&=&-\frac{1}{3} b~m_N~ g_\sigma^3 (\sigma)^{3}-\frac{1}{4} c (g_\sigma \sigma)^{4}+\frac{\xi}{4!} g_{\omega}^4 (\omega_{\mu}\omega^{\mu})^{2}  \nonumber \\
&+&\Lambda_{\omega}g_{\varrho}^{2}\boldsymbol{\varrho}_{\mu} \cdot \boldsymbol{\varrho}^{\mu} g_{\omega}^{2}\omega_{\mu}\omega^{\mu} \nonumber
\end{eqnarray}
contains the non-linear mesonic terms with parameters $b$, $c$, $\xi$, $\Lambda_{\omega}$ to take care of the high-density behavior of nuclear matter. 
The parameters $g_i$'s are the couplings of the nucleons to the meson fields $i = \sigma, \omega, \varrho$, with masses $m_i$.
Finally, the Lagrangian density for the  leptons is given as $\mathcal{L}_{leptons}= \bar{\Psi_l}\Big[\gamma^{\mu}\left(i \partial_{\mu}  
-m_l \right)\Psi_l\Big]$, where $\Psi_l~(l= e^-, \mu^-)$ denotes the lepton spinor for electrons and muons; leptons are considered non-interacting.
The equations of motion for the meson  fields are obtained from the Euler-Lagrange equations:
		\begin{eqnarray}
			{\sigma}&=& \frac{g_{\sigma}}{m_{\sigma,{\rm eff}}^{2}}\sum_{i} \rho^s_i\label{sigma}\\
			{\omega} &=&\frac{g_{\omega}}{m_{\omega,{\rm eff}}^{2}} \sum_{i} \rho_i \label{omega}\\
			{\varrho} &=&\frac{g_{\varrho}}{m_{\varrho,{\rm eff}}^{2}}\sum_{i} t_{3} \rho_i, \label{rho}
		\end{eqnarray}
 where $\rho^s_i$ and $\rho_i$ are, respectively, the scalar density and the number  density of nucleon $i$, and the  effective meson masses $m_{i,{\rm eff}}$ are defined as
 \begin{eqnarray}
   m_{\sigma,{\rm eff}}^{2}&=& m_{\sigma}^{2}+{ b ~m_N ~g_\sigma^3}{\sigma}+{c g_\sigma^4}{\sigma}^{2} \label{ms} \\ 
    m_{\omega,{\rm eff}}^{2}&=& m_{\omega}^{2}+ \frac{\xi}{3!}g_{\omega}^{4}{\omega}^{2} +2\Lambda_{\omega}g_{\varrho}^{2}g_{\omega}^{2}{\varrho}^{2}\label{mw}\\
    m_{\varrho,{\rm eff}}^{2}&=&m_{\varrho}^{2}+2\Lambda_{\omega}g_{\omega}^{2}g_{\varrho}^{2}{\omega}^{2}, \label{mr}.
 \end{eqnarray}
The meson equations are solved using the relativistic mean-field approximation for a given $\rho_n$ and $\rho_p$.
Note, that the effect of the nonlinear terms on the magnitude of the meson fields enters through the effective meson masses, $m_{i,{\rm eff}}$. The energy density of the baryons and 
leptons are given by the following expressions:
\begin{equation}
\begin{aligned}
\epsilon &= \sum_{i=n,p,e,\mu}\frac{1}{\pi^2}\int_0^{k_{Fi}} \sqrt{k^2+{m_i^*}^2}\, k^2\, dk \\
&+ \frac{1}{2}m_{\sigma}^{2}{\sigma}^{2}+\frac{1}{2}m_{\omega}^{2}{\omega}^{2}+\frac{1}{2}m_{\varrho}^{2}{\varrho}^{2}\\
&+ \frac{b}{3}(g_{\sigma}{\sigma})^{3}+\frac{c}{4}(g_{\sigma}{\sigma})^{4}+\frac{\xi}{8}(g_{\omega}{\omega})^{4} + \Lambda_{\omega}(g_{\varrho}g_{\omega}{\varrho}{\omega})^{2},
\end{aligned}
\end{equation}
   where $m_i^*=m_i-g_{\sigma} \sigma$ for protons and neutrons and $m_i^*=m_i$ for electrons and muons, and $k_{Fi}$ is the Fermi moment of particle $i$.

Once we have the energy density for a given EoS model, we can compute the chemical potential of neutron ($\mu_n$) and proton ($\mu_p$). The chemical potential of electron ($\mu_e$) and muon ($\mu_\mu$) can be computed using the condition of $\beta$-equilibrium : $\mu_n-\mu_p=\mu_e$ and $\mu_e$ = $\mu_\mu$ and the charge neutrality: $\rho_p =\rho_e + \rho_\mu$. Where $\rho_e$ and $\rho_\mu$ are the electron and muon number density.
The pressure is then determined from the thermodynamic relation:
\begin{equation}
p = \sum_{i}\mu_{i}\rho_{i}-\epsilon.
\end{equation}

\subsection{Bayesian Likelihood}

The Bayesian likelihood is a fundamental concept in Bayesian statistics, providing a way to update the probability estimate for a hypothesis as more evidence or information becomes available. In the context of Bayesian inference, the posterior distribution measures the plausibility of a set of parameters values given the observed data.

{\it Bayes' Theorem:-} Bayes' theorem is used to update the probability estimate of a hypothesis as more evidence is available. It is formulated as:
\begin{equation}
    P(\theta | {\mathcal D}) = \frac{{\mathcal L}(\mathcal D | {\theta}) P(\theta)}{P({\mathcal D})}
\end{equation}

In this equation:
\begin{itemize}
    \item \(P(\theta | {\mathcal D})\) is the posterior probability of the parameters $\theta$ given the data, \(\mathcal D\).
    \item ${\mathcal L}(\mathcal D | {\theta})$ is the likelihood of the data under the parameters (likelihood function).
    \item \(P(\theta)\) is the prior probability of the parameters $\theta$.
    \item \(P({\mathcal D})\) is the marginal likelihood or evidence, often acting as a normalizing constant.
\end{itemize}

{\it Likelihood:-} 
The likelihood function in Bayesian analysis is defined as the probability of the observed data under a specific statistical model, parameterized by a set of parameters \(\theta\). The likelihood can be computed for a set of experimental data and for the posterior distributions from the astrophysical observations as follows,

(i) Experimental data : For experimental data, D$_{expt} \pm \sigma$ having symmetric gaussian distribution of the data, the likelihood is given as,
\bea
{\mathcal L}({\mathcal D_{expt} | }\theta) &=& \frac{1}{\sqrt{2\pi\sigma^2}} exp(\frac{-(D(\theta)-D_{expt})^2}{2\sigma^2})\nonumber \\
&=& \mathcal{L}^{\rm expt}
\eea

Here \(D(\theta)\) is the model value for a given model parameter set \(\theta\). 

(ii) GW observation : For GW observations, information about EoS parameters come from the masses $m_1, m_2$ of the two binary components and the corresponding tidal deformabilities $\Lambda_1, \Lambda_2$. In this case, 
\begin{align}
    P(d_{\mathrm{GW}}|\mathrm{EoS}) = \int^{M_u}_{m_2}dm_1 \int^{m_1}_{M_l} dm_2 P(m_1,m_2|\mathrm{EoS})   \nonumber \\
    \times P(d_{\mathrm{GW}} | m_1, m_2, \Lambda_1 (m_1,\mathrm{EoS}), \Lambda_2 (m_2,\mathrm{EoS})) \nonumber \\
    =\mathcal{L}^{\rm GW}
    \label{eq:GW-evidence}
\end{align}
where P(m$|$EoS) ~\citep{Agathos_2015,Wysocki-2020,Landry_2020PhRvD.101l3007L,Biswas:2020puz} can be written as, 
\begin{equation}
    P(m|\rm{EoS}) = \left\{ \begin{matrix} \frac{1}{M_u - M_l} & \text{ iff } & M_l \leq m \leq M_u, \\ 0 & \text{ else, } & \end{matrix} \right.
\end{equation}
In our calculation we set $M_l$ = 1.36 M$_{\odot}$ and $M_u$ =1.6 M$_{\odot}$

(iii) X-ray observation(NICER) : X-ray observations give the mass and radius measurements of NS. Therefore, the corresponding evidence takes the following form,
\begin{align}
    P(d_{\rm X-ray}|\mathrm{EoS}) = \int^{M_u}_{M_l} dm P(m|\mathrm{EoS}) \nonumber \\ \times
    P(d_{\rm X-ray} | m, R (m, \mathrm{EoS})) \nonumber \\
    = \mathcal{L}^{\rm NICER}
\end{align}
 \\
Here, M$_l$ represents a mass of 1 M$_\odot$, and M$_u$ denotes the maximum mass of a neutron star according to the respective EoS.\\

\vspace{0.5cm}
The final likelihood for the three scenarios  :\\
(a) Maximum mass case (M$_{\rm J0740}$) : \\
To construct the likelihood for maximum mass we converted the X-ray mass-radius measurement of PSR J0740+6620, P(d$_{\rm X-ray}|$ m, R) to \\ P(d$_{\rm X-ray}|$ M$_{max}$)= $\mathcal{L}^{M_{\rm J0740}}$=$\int_{-\infty}^{+\infty}$ P(d$_{\rm X-ray}|$ M$_{max}$, R)dR\\

(b) M$_{\rm J0740}$ + EXPT :\\
The likelihood for this case is \\
\begin{equation}
    \mathcal{L} =\mathcal{L}^{M_{\rm J0740}}\mathcal{L}^{\rm EXPT}
\end{equation}
(c) ``ALL" : \\
The final likelihood for the ``ALL" scenario is then given by
\begin{equation}
    \mathcal{L} = 
    \mathcal{L}^{\rm EXPT}
    \mathcal{L}^{\rm GW}
    \mathcal{L}^{\rm NICER I}
    \mathcal{L}^{\rm NICER II}\,.
    \label{eq:finllhd}
\end{equation}

NICER I and NICER II  correspond to the mass-radius measurements of PSR J0030+0451 and PSR J0740+6620, respectively.
\section{Terrestrial and astrophysics constraints on EoSs}\label{data}
We use non-relativistic and relativistic mean-field models,   as briefly outlined in the previous section, to  calculate the properties of nuclear matter at several
densities and the EoSs  for the neutron star matter. The properties of nuclear matter considered are the pressure for the symmetric nuclear matter
($P_{\rm SNM}$), symmetry energy pressure ($P_{\rm sym}$) and the symmetry
energy ($e_{\rm sym}$) which are constrained  empirically over a range of
densities by the  experimental data on bulk properties of finite nuclei
such as the  nuclear masses (2), neutron skin thickness in $^{208}$Pb (1) and the
dipole polarizability (1), isobaric analog states (1) as well as the HIC data (7) spanning the density range $0.03$ - $0.32$ fm$^{-3}$, where the number of data points for a given quantity has been indicated in the parentheses. The astrophysics data considered are the mass-radius posterior distribution for PSR J0030+0451\cite{riley2019,miller2019} and PSR J0740+6620\cite{miller2021, riley2021} and posterior distribution for dimensionless tidal deformability for binary neutron star components from the GW170817 event. The EoSs are subjected to these constraints within the Bayesian framework. The  terrestrial and astrophysics data \cite{Tsang2024} considered are listed in Tab \ref{tab1}.

 \begin{table*}[]
    \centering
    \caption{\label{tab2} The median and 95$\%$ confidence interval for the nuclear matter parameters obtained for three different scenarios as labelled by $M_{\rm J0740}$, $M_{\rm J0740}$ + EXPT and ALL described in the text. The results obtained using priors subjected to the physical constraints only are also presented for comparison.}
    \begin{ruledtabular} 
    \begin{tabular}{cccccc}
  \multirow{2}{*}{NMPs(in MeV)}& \multirow{2}{*}{Models} & \multicolumn{1}{c}{Prior} & \multicolumn{1}{c}{ M$_{\rm J0740}$}  & \multicolumn{1}{c}{$M_{\rm J0740}$ + EXPT} & \multicolumn{1}{c}{ALL}\\ [1.5ex] 
  \cline{1-6}  
     \multirow{2}{*}{K$_0$}& Skyrme & 228.70$_{-59.28}^{+58.29}$&225.77$_{-42.88}^{+35.98}$ 
    &220.46$_{-35.34}^{+28.40}$  & 223.50$_{-37.60}^{+32.39}$  \\[1.5ex]
   & RMF &  239.85$_{-53.94}^{+32.48}$
&237.92$_{-42.35}^{+42.77}$&219.22$_{-30.08}^{+40.72}$  & 
     233.12$_{-45.67}^{+32.68}$ \\[1.5ex]  
  \hline 
  \multirow{2}{*}{J$_0$}& Skyrme & 33.54$_{-8.29}^{+6.46}$ &33.47$_{-8.13}^{+6.39}$ &35.41$_{-3.21}^{3.26}$  & 34.86$_{-2.98}^{+3.15}$\\[1.5ex]
       & RMF&35.63$_{-10.12}^{+26.16}$  & 29.98$_{-7.70}^{+24.15}$ &38.00$_{-1.88}^{+1.94}$  
        &38.33$_{-1.93}^{+1.87}$ \\[1.5ex]
\hline 
 \multirow{2}{*}{L$_0$}& Skyrme & 79.20$_{-60.11}^{+65.43}$
&87.60$_{-62.19}^{+56.69}$&93.65$_{-27.86}^{27.26}$ &84.50$_{-21.43}^{+27.88}$ \\[1.5ex]
& RMF& 51.79$_{-31.49}^{+84.49}$   & 50.42$_{-27.81}^{+63.10}$ &  105.57$_{-12.03}^{+7.41}$     &106.87$_{-11.80}^{+9.28}$             \\[1.5ex]
\hline
 \multirow{2}{*}{K$_{\rm sym0}$ }& Skyrme & 60.65$_{-228.77}^{+165.38}$
 &73.65$_{-190.55}^{+150.87}$  & 61.00$_{-140.56}^{+153.67}$&9.57$_{-97.09}^{+160.16}$   \\[1.5ex]
  & RMF&-153.81$_{-227.52}^{+117.38}$    & -101.18$_{-223.84}^{+87.10}$ & -12.62$_{-71.27}^{+29.59}$&-11.90$_{-65.38}^{+43.09}$  \\[1.5ex]
    \end{tabular}
    \end{ruledtabular}
\end{table*}

\begin{table*}
\centering
    \caption{\label{tab3} The median and 95$\%$ confidence interval for the radius (R), central density ($\rho_c$) and corresponding squared speed of sound (c$^2_{s_c}$) for the neutron star of mass 1.4, 2.08 M$_\odot$ and tidal deformability for 1.36 M$_\odot$  NS obtained for three different scenarios as labelled by $M_{\rm J0740}$, $M_{\rm J0740}$ + EXPT and ALL described in the text.
    }
 \begin{ruledtabular}  
\begin{tabular}{ccccc}
\multirow{2}{*}{NS properties}& \multirow{2}{*}{Models} & \multicolumn{1}{c}{ M$_{\rm J0740}$}  & \multicolumn{1}{c}{$M_{\rm J0740}$ + EXPT} & \multicolumn{1}{c}{ALL}\\ [1.5ex] 
  \cline{1-5} 
      
\multirow{2}{*}{R$_{1.4}$ (km) }& Skyrme & 13.55$_{-2.85}^{+2.09}$  & 13.65$_{-0.93}^{+1.08}$& 13.32$_{-0.85}^{+0.93}$ \\[1.5ex] 
   &     RMF   & 12.23$_{-1.01}^{+0.59}$ & 12.97$_{-0.35}^{+0.51}$  &13.12$_{-0.30}^{+0.37}$  \\[1.5ex] 
  \hline  
  
 \multirow{2}{*}{R$_{2.08}$(km)}& Skyrme& 11.88$_{-1.15}^{+1.19}$ &  11.67$_{-0.83}^{+0.79}$& 11.52$_{-0.73}^{+0.90}$  \\[1.5ex]
        & RMF   &   11.08$_{-0.71}^{+0.66}$   &11.30$_{-0.68}^{+0.65}$  &11.74$_{-0.67}^{+0.73}$           \\[1.5ex]
\hline 
 \multirow{2}{*}{$\Lambda_{1.36}$ (...)}& Skyrme & 768$_{-414}^{+658}$ & 708$_{-285}^{+261}$ & 618$_{-180}^{+268}$\\[1.5ex]
   & RMF 
   & 458$_{-193}^{+119}$  &  585$_{-125}^{+152}$  &   640$_{-109}^{+115}$            \\[1.5ex] 

\hline 
 \multirow{2}{*}{$\rho_{c,_{1.4}}$ ($\rho_0$)}& Skyrme & 2.51$_{-0.43}^{+0.78}$ & 2.66$_{-0.37}^{+0.57}$ & 2.78$_{-0.40}^{+0.47}$\\[1.5ex]
   & RMF 
   & 3.12 $_{-0.39}^{+0.45}$  &  2.92$_{-0.28}^{+0.40}$ &  2.75$_{-0.23}^{+0.31}$            \\[1.5ex] 
\hline 
 \multirow{2}{*}{$\rho_{c,_{2.08}}$ ($\rho_0$)}& Skyrme & 5.09$_{-1.08}^{+1.32}$ & 5.48$_{-1.04}^{+1.12}$ & 5.50$_{-1.18}^{+1.14}$\\[1.5ex]
   & RMF 
   & 5.62$_{-0.99}^{+1.00}$  & 5.47$_{-0.94}^{+0.99}$ & 5.00$_{-0.98}^{+1.14}$
           \\[1.5ex] 

\hline 
 \multirow{2}{*}{$c^2_{s_{c,1.4}}$}& Skyrme & 0.28$_{-0.05}^{+0.06}$ & 0.29$_{-0.04}^{+0.04}$ & 0.31$_{-0.04}^{+0.05}$\\[1.5ex]
   & RMF 
   & 0.36$_{-0.07}^{+0.09}$  & 0.32$_{-0.05}^{+0.09}$ & 0.31$_{-0.06}^{+0.11}$
           \\[1.5ex] 

\hline 
 \multirow{2}{*}{$c^2_{s_{c,2.08}}$}& Skyrme & 0.62$_{-0.11}^{+0.11}$ & 0.65$_{-0.10}^{+0.11}$ & 0.66$_{-0.11}^{+0.11}$\\[1.5ex]
   & RMF 
   & 0.57$_{-0.13}^{+0.13}$  & 0.56$_{-0.14}^{+0.14}$ & 0.50$_{-0.09}^{+0.15}$
           \\[1.5ex] 
\end{tabular}
\end{ruledtabular}
\end{table*}
\section{Results and Discussion}\label{res}
We use non-relativistic mean field model based on Skyrme type effective interaction and the relativistic mean field model derived from an effective Lagrangian to construct the EoSs as outlined in Section \ref{meth}. These EoSs are employed to calculate th NS properties and various quantities which are empirically accessible from experimental data on finite nuclei and HIC. 
The empirical values of e$_{\rm sym}$, P$_{\rm sym}$, and P$_{\rm SNM}$ for densities $\rho\leq$2$\rho_0$ have been derived from experimental data and radii, tidal deformablities and maximum mass of NS are inferred from astrophysical observations, as detailed in the preceding section.

The radius (R), the dimensionless tidal deformability ($\Lambda$) and the maximum mass of NS ($M_{\rm max}$) for a given EoS are obtained from the solutions of the Tolman–Oppenheimer–Volkoff (TOV) equations \cite{Oppenheimer:1939ne,Tolman:1939jz,Hinderer_2008}. The EoS for the outer crust up to 0.0016$\rho_0$ is taken to be the  one given by Baym-Pethick-Sutherland  \cite{Baym:1971pw}. The EoS for the inner crust corresponding to the density range $0.0016\rho_0<\rho <0.5\rho_0$  is obtained by using a polytropic form \cite{Carriere:2002bx,Patra:2022yqc}. The core EoSs are constructed using non-relativistic and relativistic mean field models as described in Sec\ref{meth}. These EoSs are subjected to (i) thermodynamic stability, (ii)  causality (speed of sound below unity), (iii) positive symmetry energy. These EoSs are further constrained within the Bayesian framework by experimental data and astrophysical observations.
The prior distributions of the parameters of these models are provided in Tabs.\ref{tabA1} and \ref{tabA2} of the appendix.

\begin{figure}
\centering
\includegraphics[width=0.5\textwidth]{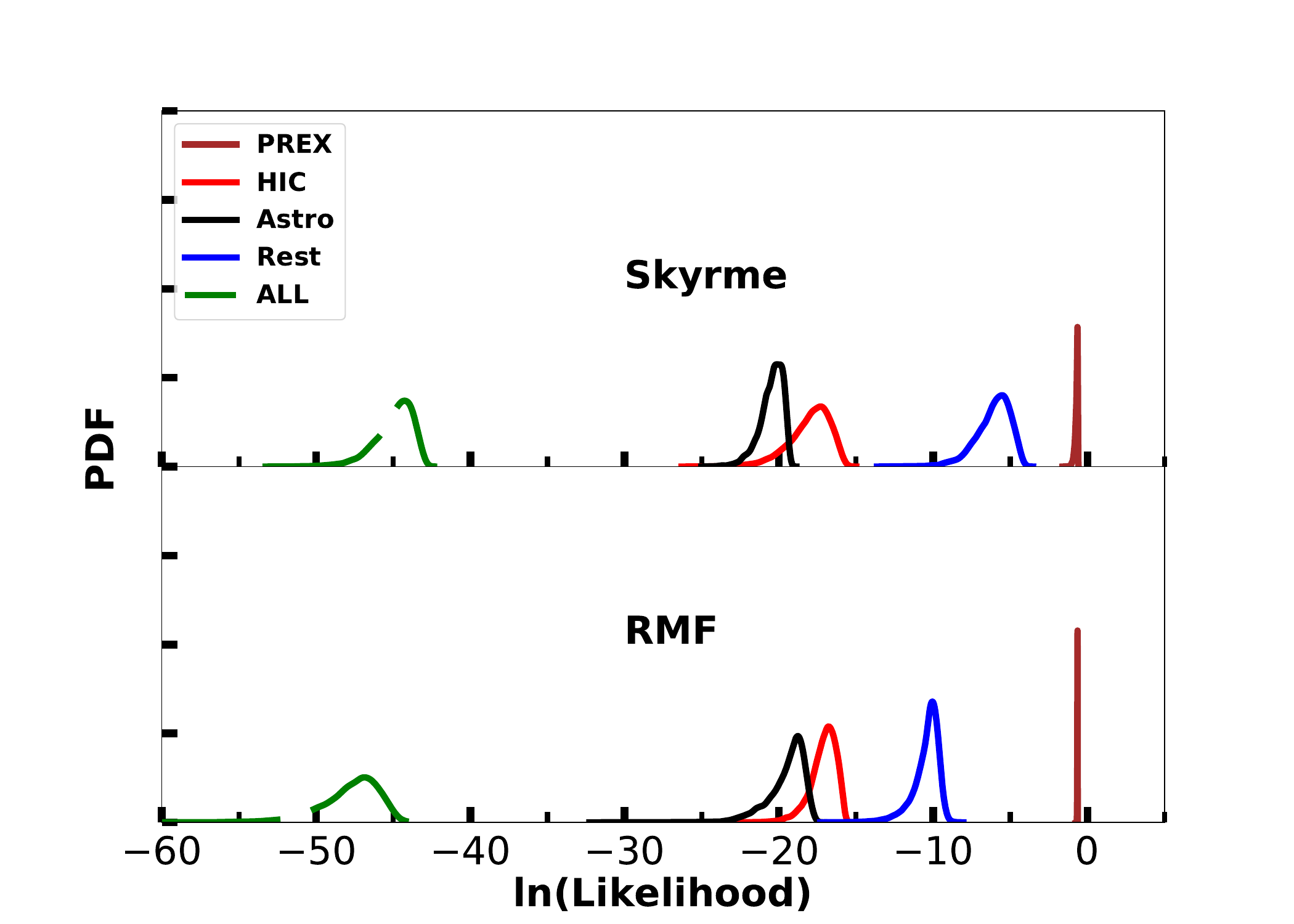}
\caption{\label{fig1}
Distributions of the value of the log likelihood for the case of all the data considered within the Skyrme (top) and RMF (bottom) models. The contributions to the logarithmic likelihood of each distinct group of data are also displayed for comparison (see text for details).}
\end{figure}
\begin{figure}
\includegraphics[width=0.5\textwidth]{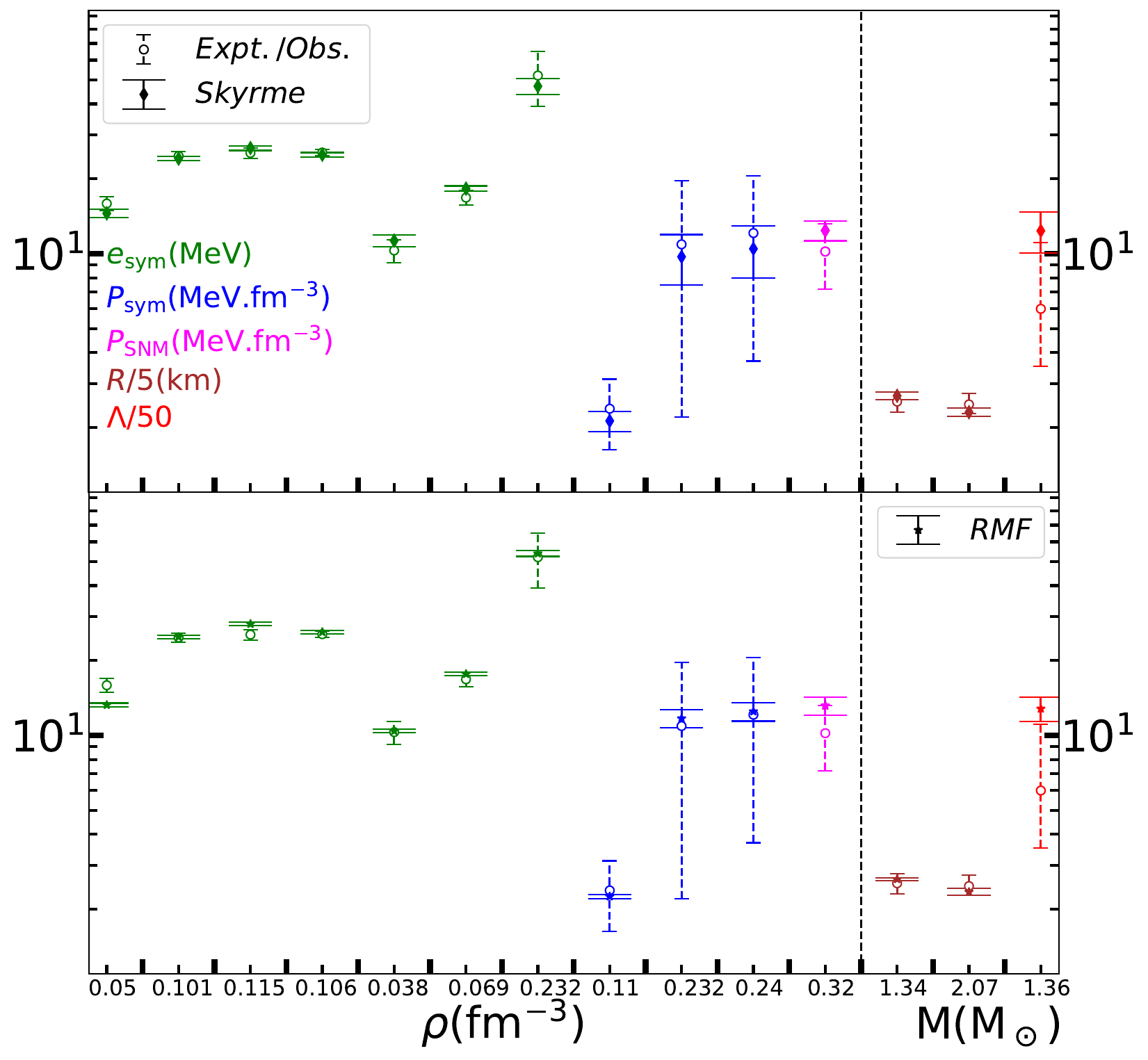}
\caption{\label{fig2}
The 68\% confidence interval for $e_{\rm sym}$, $P_{\rm sym}$, $P_{\rm SNM}$ at fixed densities (left) and radius and tidal deformability of NS (right) for the case ``ALL". The fixed values of densities and NS masses are specified along the abscissa as listed in Tab\ref{tab1}. The results are obtained using the posterior distributions of nuclear matter parameters as presented in Fig. (\ref{fig3}). The dashed vertical lines with hollow circles representing the experimental/observational data and the corresponding results from  Skyrme(top) and RMF(bottom) models are depicted by the solid vertical lines with solid circles. The scale along the ordinate only represents the numerical values, while the associated units for the presented quantities are provided in the top panel.}
\end{figure}

\begin{figure*}
\includegraphics[width=\textwidth]{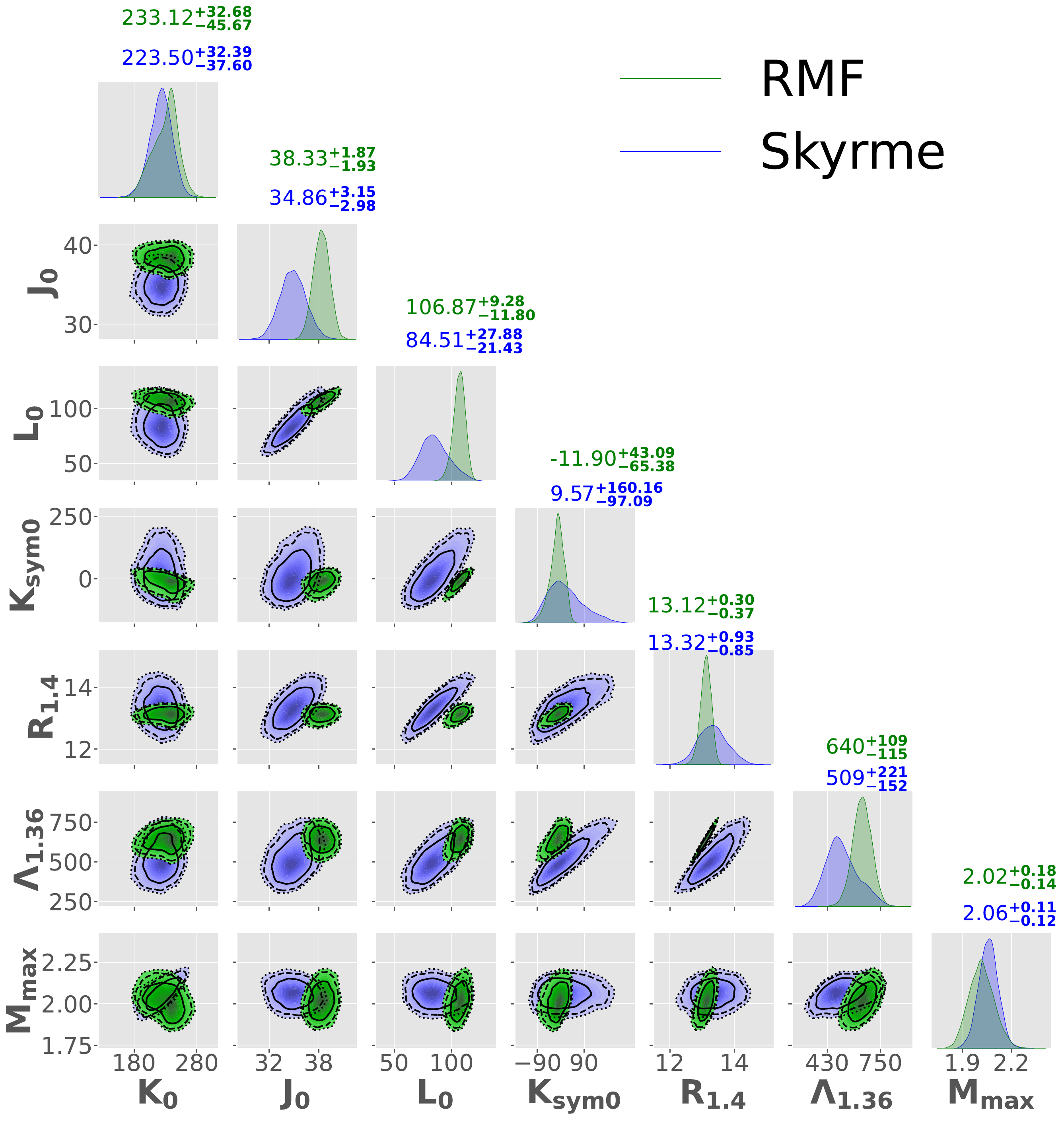}
\caption{\label{fig3}
Corner plots obtained for the Skyrme and RMF models  with the 12 experimental data and 3 astrophysical observations within the Bayesian framework. The marginalized posterior distributions of the
nuclear matter parameters and selected neutron star properties are plotted along the diagonal. The confidence ellipses for two-dimensional posterior distributions are plotted with $1\sigma$, $2\sigma$ and $3\sigma$ intervals along the off-diagonal. The errors are quoted at the 95$\%$ confidence level.
}
\end{figure*}
 In Fig.(\ref{fig1}), we display the distributions of the logarithm of the likelihood obtained for both the  EoS models considered. The curves labeled as ``ALL" (dashed) represent the results derived from employing all the data considered. Additionally, to allow a comparison, we depict contributions from various data groups. Specifically, the data groups denoted as PREX, HIC, and Astro correspond to single data from PREX-II, seven data from HIC, and five data from astrophysical observations, respectively. The group labeled as ``Rest" encompasses the remaining four data points from nuclear masses, dipole polarizability, and IAS. Looking at the figure, it is evident that the peak positions of the dashed curves  for both the models are very close to each other. The major contribution to the likelihood stems from the Astrophysical (in black) and HIC data (in red) since these are closest to the total likelihood (in dashed green) for both the models. The constraints imposed on the symmetry energy from finite nuclei data (in blue) appear to align somewhat better with the Skyrme model. The constraint on P$_{\rm sym}$ from PREX-II  seems to be fitted equally well (in brown) for both the models. In Fig. (\ref{fig2}), we display the values of $e_{\rm sym}$, $P_{\rm sym}$, and $P_{\rm SNM}$ at specific fixed densities (left), along with R$_{1.4}$, R$_{2.08}$, and $\Lambda_{1.36}$(right) within a 68$\%$ confidence interval. These values are derived from the Skyrme (top) and RMF (bottom) models within the Bayesian framework. Additionally, for comparison, we include corresponding values obtained from experimental data and astrophysical observations as used in the present work to constrain the EoS (see Tab\ref{tab1}). The data on e$_{\rm sym}$ (in green) from the nuclear masses are fitted well with the Skyrme model. The astrophysical observations related to NS properties show similar agreement for both the models considered.


To analyse systematically the implications of various data in constraining the  EoS, we perform our calculations in the following steps,
(i) Priors are updated with the posterior distribution of PSR J0740+6620 marginalised by radius. The distribution of the EoS so obtained yields  M$_{max}$ $\sim$ 2 M$_\odot$. 
(ii) Next, we further include 12 experimental data points.
(iii) All data considered in Tab\ref{tab1} are incorporated.
The outcomes for the distributions of nuclear matter parameters and neutron star properties under these three scenarios, identified as ``$M_{\rm J0740}$", ``$M_{\rm J0740}$ + EXPT", and ``ALL", are detailed in Tabs.\ref{tab2} and \ref{tab3}, respectively. 
From Tab\ref{tab2} it is clear that for the Skyrme model the K$_0$ is essentially constrained by the NS  constraint from PSR J0740+6620 , while the K$_{\rm sym0}$ is mainly constrained by the astrophysical data.  In the case of the RMF model the value of K$_0$ is influenced by both M$_{\rm J0740}$ and M$_{\rm J0740}$ + EXPT scenario. The other parameters are also significantly constrained by the experimental data  in the RMF case. Finally from the distributions of the nuclear matter parameters in the case of ALL we can conclude that the Skyrme presents a softer SNM EoS and symmetry energy compared to RMF in the vicinity of $\rho_0$. The RMF present larger values of $K_0$. This behavior  can be understood as follows: the J0740 constraint prefers a non zero value of the coupling that controls the $\omega^4$ term, and, as a consequence, at saturation the EoS has to be rather hard as discussed in \cite{Malik:2023mnx}. This will be further confirmed later when discussing the speed of sound. 

It can be seen from the Tab\ref{tab3} that in the  ``$M_{\rm J0740}$" scenario the values of radii, tidal deformabilities, central densities and corresponding squared speed of sounds for the Skyrme and RMF models are noticeable different and they practically level off when all the data are considered. It is worth noting that  although the agreement of Skyrme and RMF models with the experimental data on e$_{sym}$ are different, the EoS for the NS matter are such that the differences in the NS properties are quite close to each other. We have  also repeated all the calculations with variable transition density between 0.04 - 0.08 fm$^{-3}$ depending on the values of L$_0$ \cite{Malik:2024nva} to asses the robustness of our results. The values of radius and tidal deformability change only by less than 1$\%$ and 5$\%$, respectively.

 The corner plots in Fig.(\ref{fig3}) display nuclear matter parameters and selected NS properties under the `ALL' scenario for both Skyrme and RMF models. Along the diagonal, marginalised distributions are presented, while off-diagonal sections depict confidence ellipses. The distributions of the nuclear matter parameters for the Skyrme model are relatively wider compared to those for the RMF models. It may be emphasized that the values of nuclear matter parameters obtained from the Skyrme and RMF models differ significantly, despite being inferred from the same data set. For instance, the median value for K$_0$, J$_0$ and L$_0$ are higher in the RMF model by about 4$\%$, 10$\%$ and 25$\%$ respectively compared to the Skyrme model. This suggests that the Skyrme model yields a softer EoS for symmetric nuclear matter as well as symmetry energy around $\rho_0$. Further L$_0$ exhibits correlations with J$_0$ and K$_{\rm sym0}$ for both the models. However, the strength of correlations between the remaining pairs of nuclear matter parameters are relatively weaker. In the case of the Skyrme model, the correlations between L$_0$ and K$_{\rm sym0}$ with R$_{1.4}$ and $\Lambda_{1.36}$ appear relatively stronger compared to those in the RMF model \cite{Fortin16,Tsang:2020lmb,Thi2021,Patra:2022yqc,patra23,Imam:2023ngm,Patra:2023jbz}. 
 A possible explanation for the different behavior of RMF has already been mentioned, and it concerns the $\omega^4$ term, which has a very small influence at densities of the order of the saturation density and below, but controls the behavior of the EoS at high densities. This means that the low density and high density behavior in RMF models are quite decoupled.
 It may be pointed out that R$_{1.4}$ is better constrained for the RMF model. This may be associated to the narrow distribution of L$_0$.

\begin{figure} [ht!]
\includegraphics[width=0.5\textwidth]{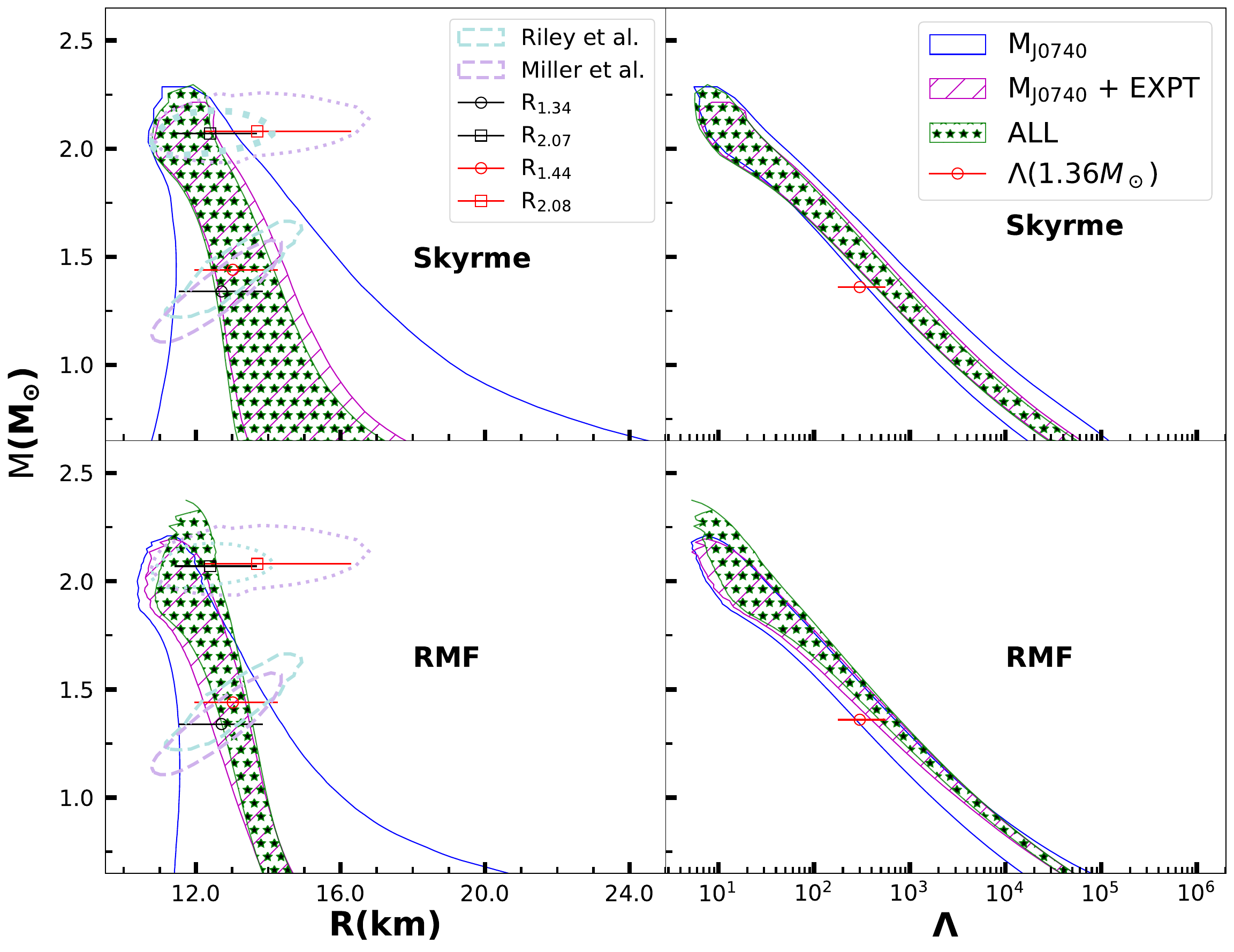}
\caption{\label{fig4}
The $95\%$ confidence intervals for the radius and tidal deformability as a function of neutron star mass evaluated using the posterior distributions. The astrophysical observations used in the Bayesian framework are also shown. The error bars in black and red colors in the left panels indicate the constraints from two different analysis for NS radii for each of the pulsar PSR J0030+0451 and PSR J0740+6620 with masses $\sim$ 1.4 and 2.0 M$_\odot$, respectively. Whereas, the error bars in red in the right panel indicate the tidal deformability constraint from the GW170817 event. 
}
\end{figure}

We examine the impact of three different scenarios on the distributions of radius, tidal deformability of NS and the underlying EoSs. In Fig.(\ref{fig4}), the $95\%$ confidence intervals display the NS mass-radius and mass-tidal deformability relationships. Results from both the Skyrme and the RMF models exhibit similar trends. The incorporation of experimental data (magenta band) significantly reduces the spread in NS properties at a given mass compared to the scenario involving only the maximum mass constraint (empty blue band). The values of R$_{1.4}$, R$_{2.08}$, and $\Lambda_{1.36}$ are already reasonably constrained by $M_{\rm J0740}$ + EXPT scenario, aligning well with current astrophysical observations. Interestingly, the inclusion of constraints from astrophysical observations (green band) do not significantly alter the NS properties. This suggests a need for more precise observations to further refine and constrain the underlying EoS.

\begin{figure} [ht!]
\includegraphics[width=0.5\textwidth]{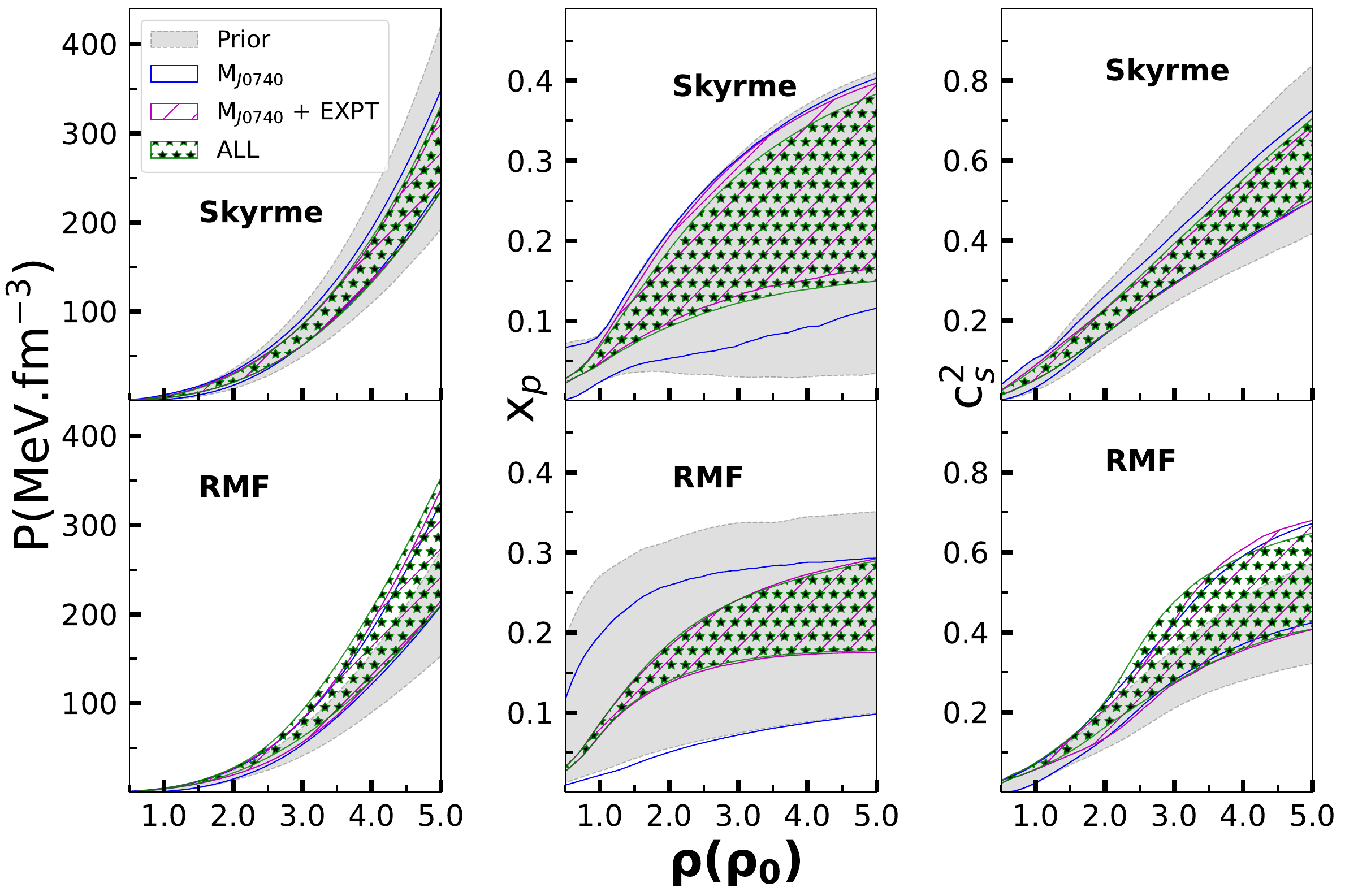}
\caption{\label{fig5} 
The  $95\%$ confidence intervals for the posterior distributions  of the pressure (P), proton fraction (x$_p$) and squared speed of sound (c$_s^2$) for the neutron star matter  as a function of baryon density for the Skyrme (top) and RMF (bottom) mean field  models corresponding to different scenarios as indicated (see text).}
\end{figure}

In Fig.(\ref{fig5}), the $95\%$ confidence intervals for the posterior distributions  of the pressure (P), proton fraction (x$_p$) and square of speed of sound (c$_s^2$) for the neutron star matter represent the updated EoSs achieved by sequentially incorporating the constraints. For the sake of comparison the EoSs corresponding to the original priors (grey band) are also depicted. The J0740 mass constraint significantly constrains the EoSs, proton fraction  as well as squared speed of sound  (empty blue band) which further narrow down with the inclusion of experimental data (magenta band). The introduction of constraints from astrophysical observations does not notably impact the EoSs, proton fraction and squared speed of sound. It is interesting to note the different predictions for the speed of sound and the proton fraction in the two approaches: while the squared speed of sound seems to saturate for RMF models around 0.6, Skyrme models show a linear increase with density, taking values that can go beyond 0.7 at 5$\rho_0$; the spread on $x_p$ is much larger for Skyrme models, taking values up to 0.4 or more, while RMF do not go beyond 0.3. In Table \ref{tab3}, we give the central squared speed of sound  and density of 1.4 M$_\odot$ and 2.08 M$_\odot$ stars. After applying all constraints the squared speed of sound takes similar values for 1.4 M$_\odot$ stars in both descriptions, but Skyrme models predict a quite larger value for 2.08 M$_\odot$ stars, the medians being 0.66 for Skyrme models and 0.5 for RMF models. Concerning the central densities, Skyrme models predict  slightly  larger densities, and, in particular, for the 2.08 M$_\odot$ stars a larger compactness because a smaller radius is also predicted. 

\begin{figure} 
\includegraphics[width=0.5\textwidth]{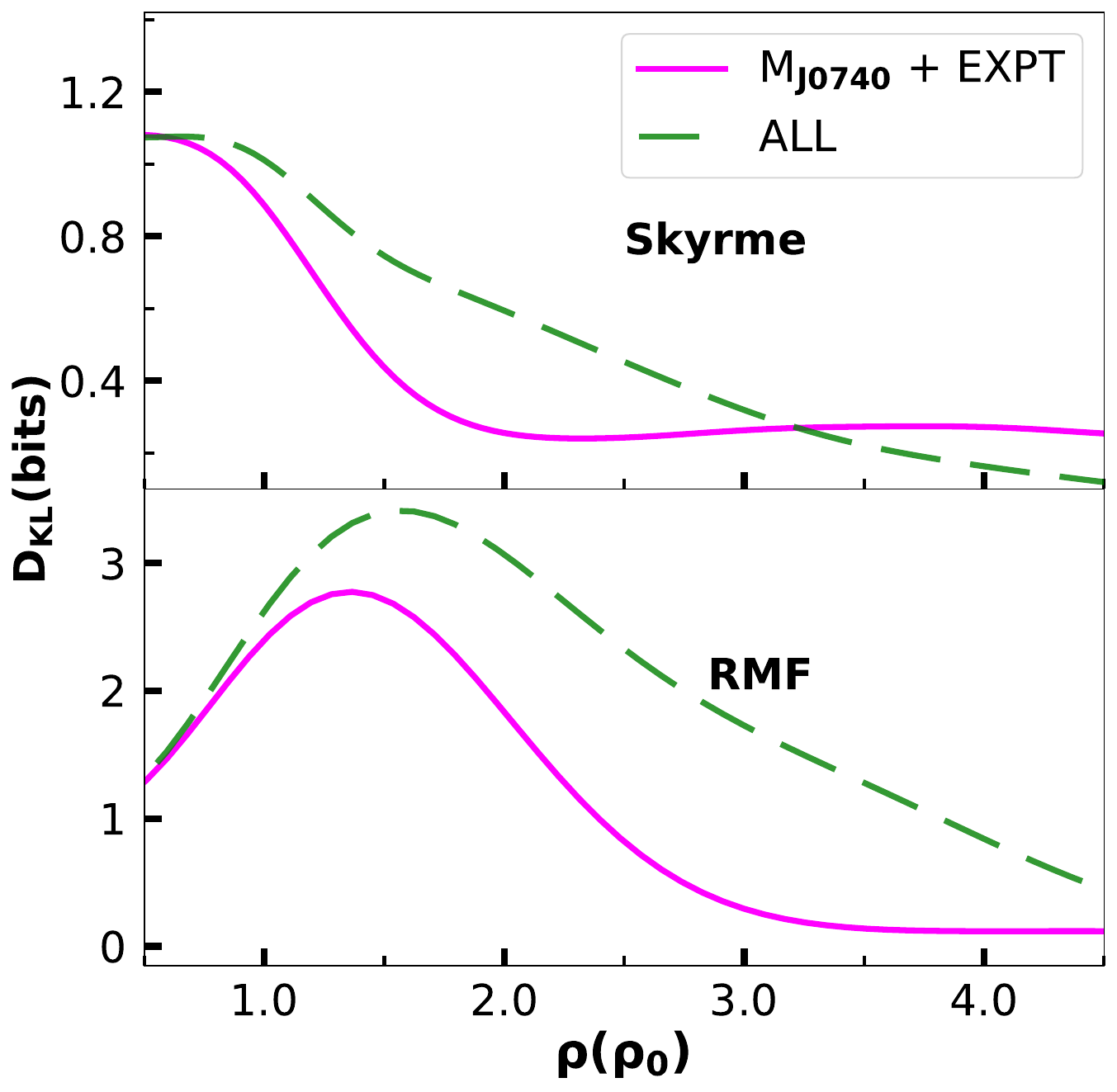}
\caption{\label{fig6}
 The variations of Kullback–Leibler divergence (D$_{KL}$) as a function of baryon density. The values of D$_{KL}$ shown by solid (dashed) line is  obtained for the scenario M$_{\rm J0740}$ + EXPT (ALL) with respect M$_{\rm J0740}$ scenario. }
\end{figure}

We also perform a quantitative comparison between the EoSs from different scenarios through the Kullback-Liebler divergence (D$_{KL}$). The D$_{KL}$ comparing distributions P and Q is defined as :\\
D$_{KL}$(P$\mid\mid$Q) = $\int_{-\infty}^{+\infty}$ P(x)log$_2$($\frac{P(x)}{Q(x)}$)dx.
We evaluate the D$_{KL}$ as a function of baryon density for two distributions of the EoSs corresponding to two different scenarios. In Fig.(\ref{fig6}), the solid (dashed ) lines represent the value of D$_{KL}$ obtained for the distributions of EoSs  for ``$M_{\rm J0740}$ + EXPT" ("ALL") scenario with respect to the ``$M_{\rm J0740}$" scenario. The values of D$_{KL}$ for the solid lines indicate that the experimental data provides tighter constraints on the EoS in the vicinity of $\rho_0$. The proximity between solid and dashed lines indicates that including astrophysical observations does not impose tighter constraints beyond those set by the experimental data. However, the differences between the solid and dashed curves around 2-3$\rho_0$ in both the models suggests that astrophysical observations on NS radii and tidal deformability marginally constrain the EoS within this density range.

Finally, we would like to compare our results on the radius of the 1.4$M_\odot$ star with other results existing in the literature.  In Ref.\cite{Pang:2022rzc}, the authors made an estimation of the radius of a 1.4$M_\odot$ star, $R_{1.4}=11.98^{+0.35}_{-0.40}$~km, using an extension of the  Nuclear-physics and Multi-Messenger Astrophysics framework NMMA, which includes data from gravitational-wave signal from the GW170817 event, the kilonova AT2017gfo and the gamma-ray burst GRB170817A afterglow, together with nuclear-physics constraints at low densities and X-ray and radio observations of isolated neutron stars. This predicted value is in agreement with several other predictions listed in Tab\ref{tab1} of Ref.\cite{Pang:2022rzc} which have been obtained essentially from observational data and nuclear-physics constraints at low densities for $\chi$EFT calculations for neutron matter, except for the study \cite{Huth_2021}  where data from HIC \cite{Russotto16,LeFevre:2015paj, Danielewicz:2002pu} has also been considered. In Ref.\cite{Huth_2021} the predicted radius of a 1.4$M_\odot$ NS is 12.01$^{+0.78}_{-0.77}$ km considering Astro and HIC constraints, however taking only the HIC constraints the radius is larger, 12.06$^{+1.13}_{-1.18}$  km. The quite large radius predicted in our study 13.32$_{-0.85}^{+0.93}$  km (Skyrme) and  13.12$_{-0.30}^{+0.37}$ km (RMF) is compatible with the prediction from \cite{Huth2021} from the HIC data. Note, however, that in our study we have considered a larger set of HIC and experimental nuclear data. In Ref.\cite{Tsang2024} the R$_{1.4}$ = 12.9$_{-0.5}^{+0.4}$ km has been obtained using a diverse set of data from finite nuclei, HIC, GW170817 and NICER which compare well with R$_{1.4}$ $\sim$ 13 km as obtained in the present work (see Tab\ref{tab3}). These results suggest that 
lower values of R$_{1.4}$ obtained in \cite{Pang:2022rzc} and references therein may be due to the constraints from $\chi$EFT for pure neutron matter at low density. In a recent study \cite{Breschi:2024qlc} a combined analysis has been performed using GW170817, kilonova and NICER data which predict radius of 1.4 M$_\odot$ NS 12.30$_{-0.56}^{+0.81}$ km and 13.20$_{-0.91}^{+0.90}$ km for two different hot spot models used in the NICER analysis.

\section{Conclusions}\label{conc}
We have performed a systematic investigation to assess the implications of a comprehensive set of data, encompassing nuclear physics experiments and astrophysical observations concerning neutron star properties, on the EoSs obtained within the non-relativistic and relativistic mean field models. The non-relativistic model relies on a Skyrme-type effective interaction, while the relativistic mean field model is derived from an effective Lagrangian which  incorporates nonlinear terms for $\sigma$, $\omega$, and $\rho$ mesons. The experimental data set includes bulk properties from finite nuclei and heavy-ion collisions, contributing with information on the symmetry energy, the symmetry energy pressure, and the pressure for symmetric nuclear matter up to twice the saturation density.  Additionally, the astrophysics data considered are the mass-radius posterior distribution for PSR J0030+0451 and PSR J0740+6620 and posterior distribution for dimensionless tidal deformability for binary neutron star components from the GW170817 event. We have employed a Bayesian framework, to investigate the constraints imposed by these data on nuclear matter properties. Three scenarios are explored: (i) priors updated with the PSR J0740+6620 mass constraint only, (ii) PSR J0740+6620 mass constraint  plus experimental data, and (iii) all the data considered.

Our determined values largely align within 68$\%$ confidence intervals with most of the fitted data/observations (see Fig.\ref{fig2}). The RMF model tends to overestimate the symmetry energy at densities of 0.05 and 0.115 fm$^{-3}$ derived from iso-vector dipole resonance and nuclear masses, respectively. The dimensionless tidal deformability from the Skyrme model aligns slightly better with the corresponding astrophysical observations for a neutron star of 1.36 M$_\odot$.
The underlying EoSs initially constrained by the PSR J0740+6620 mass, narrow further with the inclusion of experimental data. Interestingly, the impact on the EoSs by further inclusion of astrophysical observations is only marginal. This is especially evident in the distributions of neutron star mass-radius and mass-tidal deformability. These trends are also quantitatively reflected from the values of KL divergence obtained by comparing the distributions of the EoS from different scenarios. It implies that more precise observations are  required to confine the EoSs in narrower bounds. Finally, we remark that there is a significant discrepancy in the values of nuclear matter parameters obtained from the Skyrme and RMF models, despite both originating from the same data set. In particular,  as alluded above,  Skyrme forces predict a softer symmetric nuclear matter EoS as well as softer symmetry energy around saturation density. Note, however, that although the Skyrme forces predict a softer EoS at saturation density, they are also the ones that predict larger values for the speed of sound and the proton fraction in the center of $\sim 2$ M$_\odot$ stars. It is also observed that the value of R$_{1.4}\sim$ 12 km in the previous analysis may be due to the low density constraints from $\chi$EFT for pure neutron matter.

\section{Acknowledgements} 
TM and CP would like to appreciate the support of national funds from FCT (Fundação para a Ciência e a Tecnologia, I.P, Portugal) under projects  UIDP/\-04564/\-2020 and No. UIDB/\-04564/\-2020, respectively with DOI identifiers 10.54499/UIDB/04564/2020 and 10.54499/UIDP/04564/2020,  and Project  2022.06460.PTDC with the  DOI identifier 10.54499/2022.06460.PTDC. They also like to acknowledge the Laboratory for Advanced Computing at the University of Coimbra for providing {HPC} resources that have contributed to the research results reported within this paper, URL: \hyperlink{https://www.uc.pt/lca}{https://www.uc.pt/lca}. BKA is thankful to the University of Coimbra, CFisUC, for their partial support during the visit, where the current work was initiated. BKA also acknowledges partial support from the SERB, Department of Science and Technology, Government of India with  CRG/2021/000101. The authors sincerely acknowledge the usage of the analysis software {\tt BILBY} \cite{Ashton2019,Bilby_ref}.

\bibliographystyle{apsrev4-2}


\pagebreak
\setcounter{equation}{0}
\setcounter{figure}{0}
\setcounter{table}{0}
\setcounter{page}{1}
\makeatletter
\renewcommand{\theequation}{A\arabic{equation}}
\renewcommand{\thefigure}{A\arabic{figure}}
\renewcommand{\thetable}{A\arabic{table}}

\widetext
\section*{Appendix}
\begin{table}[H]
\caption{ Prior distributions of the NMPs for the Skyrme model. Here $\rho_0$, e$_0$ and K$_0$ are assumed as gaussian (G) distribution with median $\mu$ and standard deviation $\sigma$. Uniform (U) priors are used for the rest of the parameters with a minimum min and maximum max. The NMPs are listed in the units of MeV. Effective mass m$^*$ has the unit of nucleon mass, m$_N$.} \label{tabA1}
  \centering
  \begin{ruledtabular}  
  \begin{tabular}{cccccccc}
 &{$\rho_0$}(G) & {e$_0$}(G) & {K$_0$}(G) & {J$_0$}(U) &{L$_0$}(U) & {K$_{\rm sym,0}$}(U)  & m$^*$(U)\\ [1.3ex]
 \hline
 min/$\mu$ & 0.16 & -16 & 230 & 24.7 & -11.4 & -328.5  & 0.5 \\[1.3ex] 
 max/$\sigma$ & 0.005 & 0.3 & 30 & 40.3 & 149.4 & 237.9 & 1\\[1.3ex] 
  \end{tabular}
  \end{ruledtabular}
\end{table}



\begin{table}[H]
\caption{The uniform prior is considered for the parameters of the RMF models (NL). Specifically, B and C are \(b \times 10^3\) and \(c \times 10^3\), respectively. It follows that "b" and "c" are going into Lagrangian density. The 'min' and 'max' entries denote the minimum and maximum values of the uniform distribution. All the parameters are dimensionless. Additional row for effective priors in NMP is included (90\% CI). It is to be noted to calculate effective priors of NMPs in the RMF model only the e$_0$ and $\rho_0$ constrains are implemented.}\label{tabA2}

\setlength{\tabcolsep}{4pt} 
\renewcommand{\arraystretch}{1.6}
\begin{tabular}{lccccccc}
\toprule
 & \(g_{\sigma}\) & \(g_{\omega}\) & \(g_{\varrho}\) & B & C & \(\xi\) & \(\Lambda_\omega\) \\
\hline 
min & 5.5 & 5.5 & 5.5 & 0.5 & -10.0 & 0.0 & 0 \\
max & 15.5 & 15.5 & 20.5 & 10.0 & 10.0 & 0.04 & 0.12 \\
\hline
NMP & \(\rho_0=0.16\pm0.01\) & \(e_0=-16\pm0.03\) & \(K_0=240_{-30}^{+25}\) & \(Q_0=-565_{-50}^{+60}\) & \(J_0=36_{-12}^{+36}\) & \(L_0 =55_{-30}^{+85}\) & \(K_{\rm sym0}=-166_{-300}^{+120}\) \\
\hline 
\end{tabular}
\end{table}

\end{document}